\documentclass[
    ,final            
  ]
  {aipproc}

\layoutstyle{6x9}

\def\mevc  {\ifmmode {\rm MeV}/c \else MeV$/c$\fi}
\def\mevcc {\ifmmode {\rm MeV}/c^2 \else MeV$/c^2$\fi}
\def\gevc  {\ifmmode {\rm GeV}/c \else GeV$/c$\fi}
\def\gevcc {\ifmmode {\rm GeV}/c^2 \else GeV$/c^2$\fi}
\def\ra    {\rightarrow}
\def\Bz    {\ensuremath{B^0}}
\def\Bs    {\ensuremath{B_s^0}}
\def\Bsb   {\ensuremath{\bar{B}_s^0}}
\def\BsH   {\ensuremath{B_s^H}}
\def\BsL   {\ensuremath{B_s^L}}
\def\BsJPsiPhi  {\ensuremath{\Bs\ra J/\psi\phi}}
\def\dGs   {\ensuremath{\Delta\Gamma_s}}
\def\dMs   {\ensuremath{\Delta m_s}}

\def\bsj   {\ensuremath{\beta_s^{J/\psi\phi}}}
\def\bssm  {\ensuremath{\beta_s^{SM}}}
\def\psnp  {\ensuremath{\phi_s^{NP}}}
\def\Sb    {\ensuremath{\Sigma_b}}
\def\Xib   {\ensuremath{\Xi_b^-}}
\def\Omb   {\ensuremath{\Omega_b^-}}
\def\Lb    {\ensuremath{\Lambda_b^0}}

\begin{document}

\title{Physics of \Bs~Mesons and Bottom Baryons}


\author{Manfred Paulini}{
  address={Carnegie Mellon University, Pittsburgh, PA 15213, USA}
}

\begin{abstract}
  We discuss the physics of \Bs~mesons focusing on $CP$~violation in
  \BsJPsiPhi~decays at the Tevatron. We summarize measurements of the
  properties of bottom baryons at the Tevatron including the \Sb~states
  and the \Xib~baryon. We also discuss the discovery of the \Omb~baryon.
\end{abstract}

\maketitle


\section{Introduction}
The past decade has seen an overwhelming amount of exciting heavy
flavour physics results~\cite{Amsler:2008zzb} from the
$e^+e^-$~$B$~factory experiments BaBar and Belle as well as the CDF and
D0~experiments operating at the Tevatron $p\bar p$~collider.  In many
cases, the measurements performed at the Tevatron Collider are
complementary to those at the $B$~factories. In particular, all
$B$~hadron states are produced at the Tevatron. Besides the neutral
$B^0$ and the charged $B^+$ which are the only products at the
$\Upsilon(4S)$~resonance, the Tevatron is also a source of $B$~mesons
containing $s$- or $c$-quark: \Bs\ and $B_c^+$. In addition, baryons
containing bottom quarks such as the \Lb, \Xib, $\Sb^-$ or \Omb~baryons
are produced at the Tevatron.

Why do we study $B$~hadron states?  In analogy to the hydrogen atom
which consists of a heavy nucleus in form of the proton surrounded by a
light electron, a $B$~hadron consists of a heavy bottom quark surrounded
either by a light anti-quark, to form a $B$~meson or a di-quark pair, to
form a bottom baryon. The interaction between the $b$~quark and the
other quark(s) in a $B$~hadron is based on the strong interaction or
Quantum Chromodynamics (QCD) while the interaction between proton and
electron is based on the electromagnetic Coulomb interaction and
described by Quantum Electrodynamics in its ultimate form.  Heavy quark
hadrons are often called the hydrogen atom of QCD. The study of
$B$~hadron states is thus the study of (non-perturbative) QCD, providing
sensitive tests of all aspects of QCD, including lattice gauge
calculations. In addition, the study of the Cabibbo-Kobayashi-Maskawa
(CKM) mechanism which governs quark transitions allows for precision
tests of the standard model (SM) and the search for physics beyond the
SM through measurements of loop processes in which non-SM particles can
contribute.

After a successful 1992-1996 Run\,I data taking period (for a review of
$B$~physics results from e.g.~CDF in Run\,I see
Ref.~\cite{Paulini:1999px}), the Fermilab Tevatron operates in Run\,II
at a centre-of-mass energy of 1.96~TeV with a bunch crossing time of
396~ns generated by $36\times36$ $p\bar p$ bunches. The initial Tevatron
luminosity steadily increased from 2002 to 2009 with a present peak
luminosity of $35\cdot 10^{31}$~cm$^{-2}$s$^{-1}$ reached in 2009.  The
total integrated luminosity delivered by the Tevatron to CDF and D0 at
the time of this conference is $\sim\!6.5$~fb$^{-1}$ with about
$5.5$~fb$^{-1}$ recorded to tape by each collider experiment.  However,
most results presented in this review use about 1-4~fb$^{-1}$ of data.

\section{Physics of  \Bs~Mesons}

In the neutral \Bs~system there exist two flavour eigenstates, the 
$\Bs=|\bar b s\,\rangle$ and $\Bsb=|b\bar s\,\rangle$. 
The mass eigenstates \BsH\ and \BsL\ are admixtures of the flavour
eigenstates \Bs\ and \Bsb:
\begin{equation}
|\BsH\,\rangle = p\,|\Bs\,\rangle - q\,|\Bsb\,\rangle,\quad\quad
|\BsL\,\rangle = p\,|\Bs\,\rangle + q\,|\Bsb\,\rangle,\quad {\rm{with}}\ \
\frac{q}{p}=\frac{V_{tb}^*V_{ts}}{V_{tb}V_{ts}^*}.
\end{equation}
The fact that the mass eigenstates are not the same as the flavour
states gives rise to oscillations between the \Bs\ and \Bsb~states with
a frequency proportional to the mass difference of the mass eigenstates,
$\dMs=m_H-m_L$. In the SM particle-antiparticle oscillations are
explained in terms of second-order weak processes involving virtual
massive particles that provide a transition amplitude between the \Bs\
and \Bsb~states. The decay width difference between the mass eigenstates
$\dGs=\Gamma_L-\Gamma_H\sim2\,|\Gamma_{12}|\cos\phi_s$ is related to the
$CP$~phase $\phi_s=\rm{arg}(-M_{12}/\Gamma_{12})$ where $M_{12}$ and
$\Gamma_{12}$ are the off-diagonal elements of the mass and decay
matrix. 
The decay \BsJPsiPhi\ is the transition of the spin-0 pseudo-scalar \Bs\
into two spin-1 vector particles. The orbital angular momenta of the
vector mesons, $J/\psi$ and $\phi$, can be used to distinguish the
$CP$~even $S$-wave ($L=0$) and $D$-wave ($L=2$) final states from the
$CP$~odd $P$-wave ($L=1$) final state. Such an
angular decomposition reveals that the decay is dominated by the
$CP$~even state.

\subsection{$CP$~Violation in \BsJPsiPhi} 

In analogy to measurements of the time dependent $CP$~asymmetry in
neutral \Bz~decays into e.g.~$\Bz\ra J/\psi K_S^0$ accessing the $CP$
violating phase $\sin(2\beta)$ which arises through the interference
between decay and mixing, the application of flavour tagging to
\BsJPsiPhi\ events measures the corresponding phase in \Bs~decays.  This
phase, which is responsible for $CP$~violation in \BsJPsiPhi\ in the
standard model, is in analogy to the phase $\sin(2\beta)$ called
$\sin(2\bssm)$ and is defined through the CKM matrix elements as 
$\bssm={\rm arg}(-V_{ts}V_{tb}^*/V_{cs}V_{cb}^*)$.  
In the context of the SM, this phase is expected to be small
($2\bssm\sim0.04$) and its measurement is currently beyond the
experimental reach of the Tevatron.  However, new physics may contribute
significantly larger values to the $CP$~violating phase in
\BsJPsiPhi~decays~\cite{Lenz:2006hd,Ligeti:2006pm,Hou:2006mx}.  In this
case, the observed $CP$~phase would be modified by a phase \psnp\ due to
new physics processes, and can be expressed as $2\bsj=2\bssm-\psnp$.
The current interest in measuring $CP$~violation in \BsJPsiPhi\ is
therefore in searching for enhanced $CP$~violation through new physics
processes.

\begin{figure}[tb]
\centering
\includegraphics[height=0.17\textheight]{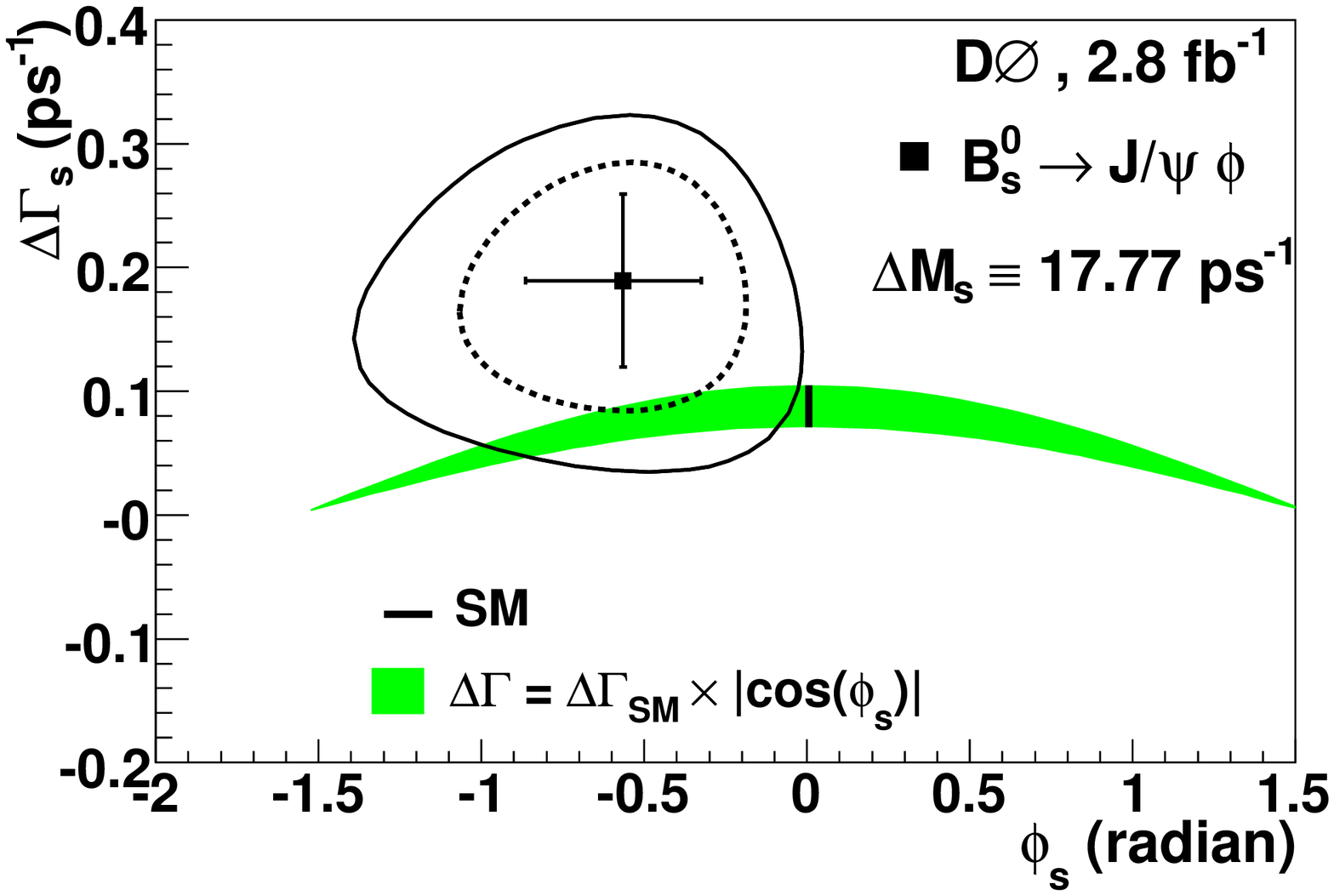}
\includegraphics[height=0.18\textheight]{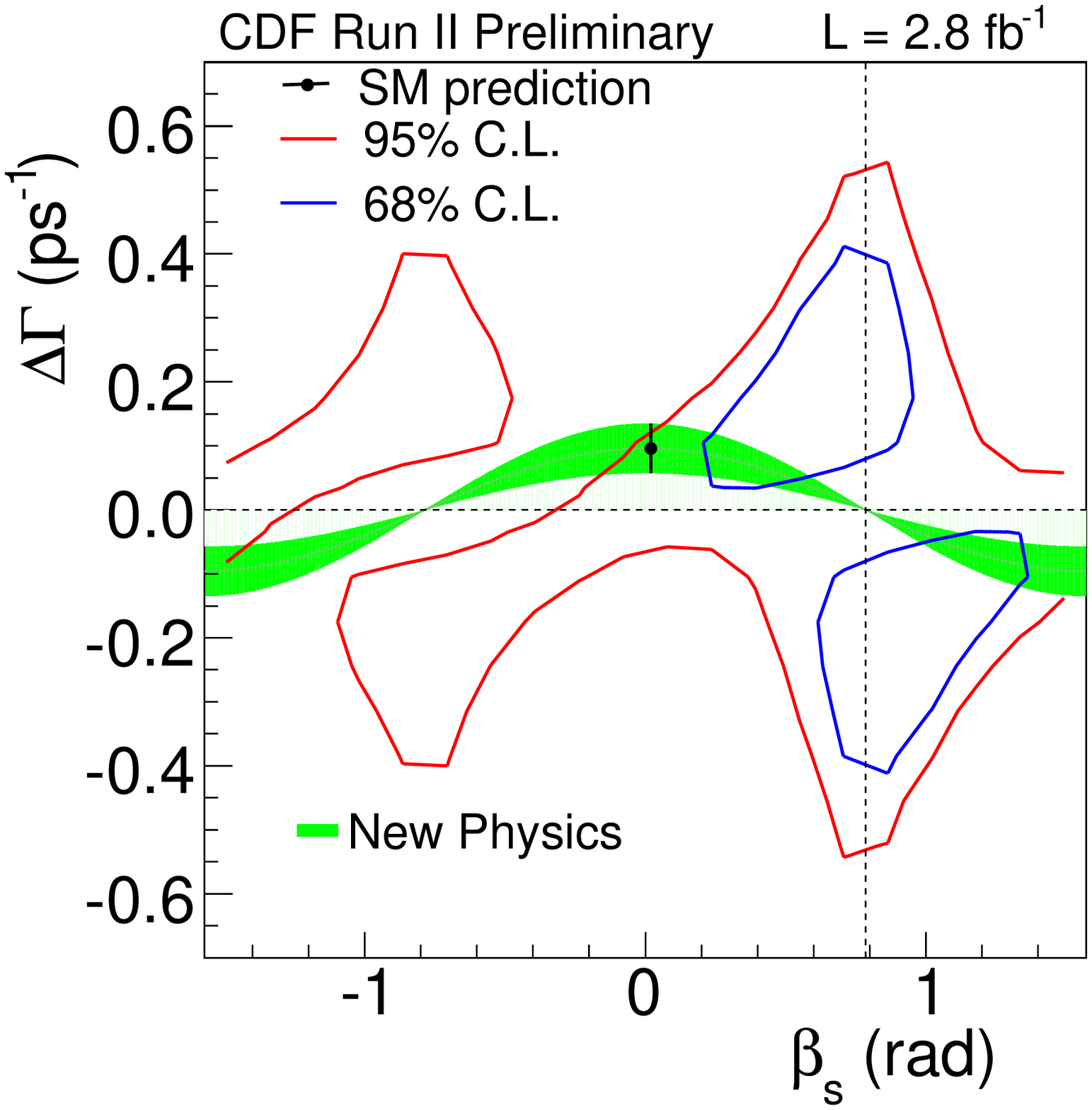}
\includegraphics[height=0.19\textheight]{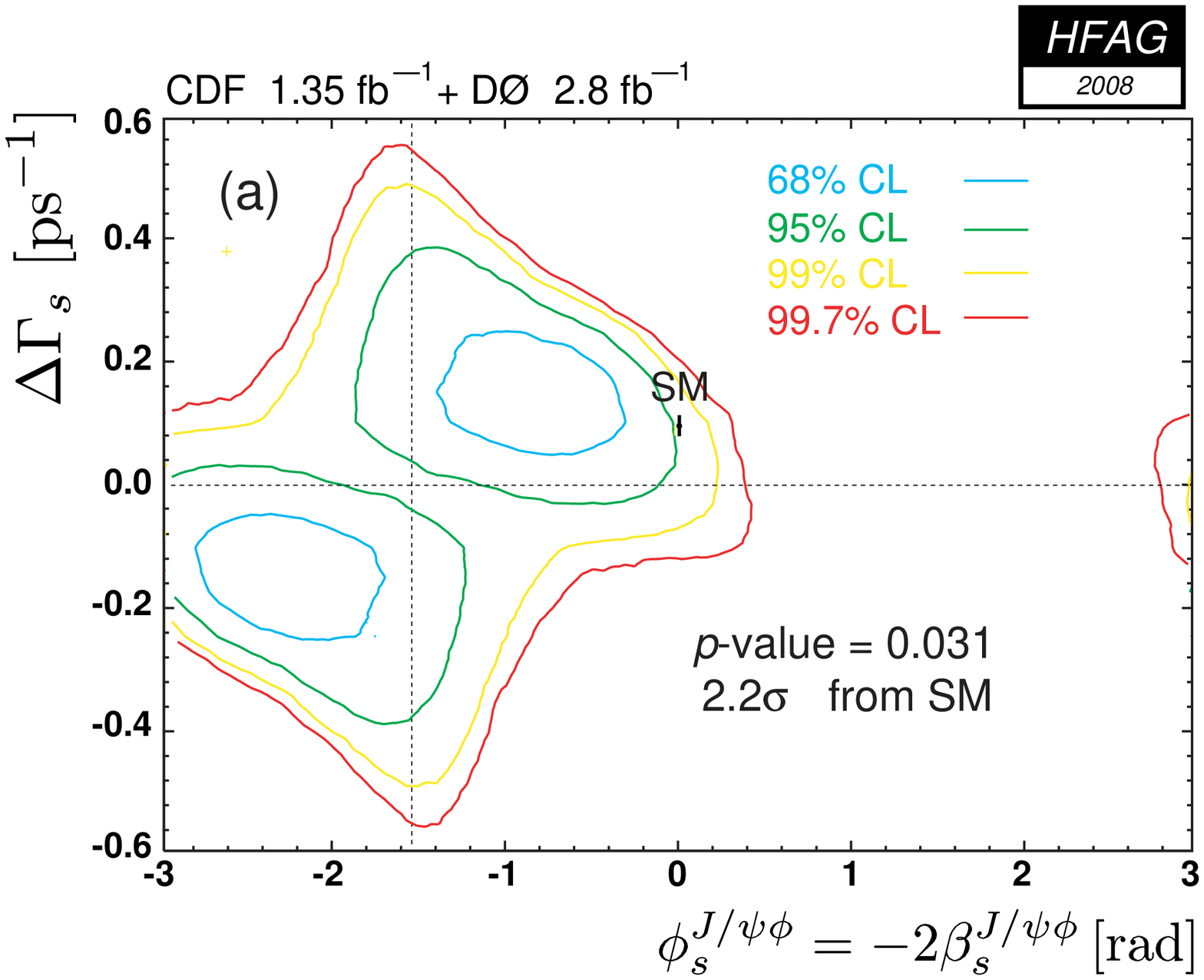}
\put(-395,90){\bf (a)}
\put(-196,91){\bf (b)}
\put(-123,92){\bf (c)}
\caption{Confidence level regions in plane of \dGs\ versus \bsj\ in
  flavour-tagged \BsJPsiPhi\ analysis from (a) the 2.8~fb$^{-1}$ result
  from D0, (b) the preliminary 2.8~fb$^{-1}$ result from CDF and (c)
  the combination of CDF and D0 likelihoods. Note the transformation
  $2\bsj=-\phi_s^{J/\psi\phi}$.}
\label{fig:Bscontour}
\end{figure}

At the time of this conference, both Tevatron experiments have presented
tagged, time dependent angular analyses of \BsJPsiPhi~decays.  Due to
the non-parabolic behaviour of the log-likelihood function, no
meaningful point estimates for \bsj\ can be quoted and both experiments
construct their results as confidence level regions in the plane of
\dGs\ versus \bsj. The D0 result~\cite{Abazov:2008fj} based on
2.8~fb$^{-1}$ of data is shown in Figure~\ref{fig:Bscontour}(a) while a
published result from CDF based on 1.35~fb$^{-1}$ of
data~\cite{Aaltonen:2007he} exists. A preliminary update from CDF with
2.8~fb$^{-1}$ of data is displayed in Figure~\ref{fig:Bscontour}(b).
Both experiments observe a mild inconsistency with the SM prediction
$2\bssm\sim0.04$. Interestingly, the CDF and D0 inconsistencies with the
standard model both point in the same direction. Assuming the SM
prediction, CDF quotes a probability of 7\% to observe a likelihood
ratio equal or higher than the one observed in data which corresponds to
about $1.8\,\sigma$. Using constraints on the strong phases, D0 finds a
$p$-value of 6.6\% corresponding to a $1.8\,\sigma$ inconsistency with
the SM hypothesis~\cite{Abazov:2008fj}. The combination of the results
from Refs.~\cite{Abazov:2008fj} and \cite{Aaltonen:2007he} is shown in
Figure~\ref{fig:Bscontour}(c), restricting \bsj\ to the interval
$[0.14,0.73]\cup[0.83,1.42]$ at 90\% confidence level. The
consistency of the combined result with the SM gives a $p$-value of
3.1\% corresponding to a $2.2\,\sigma$ discrepancy with the SM
prediction.

\section{Physics of Bottom Baryons}

\begin{figure}[tb]
\centering
\includegraphics[height=0.27\textheight]{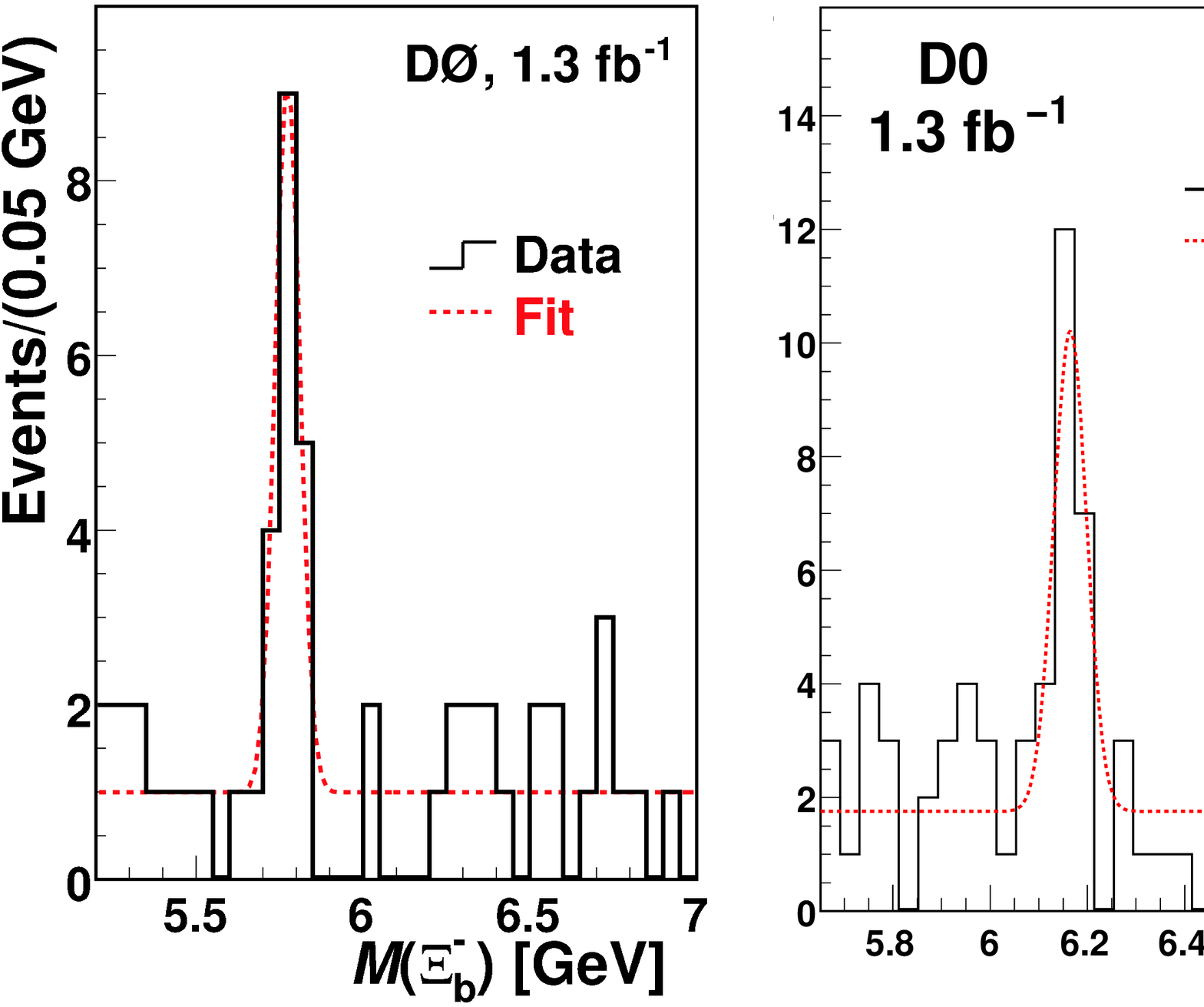}
\includegraphics[height=0.27\textheight]{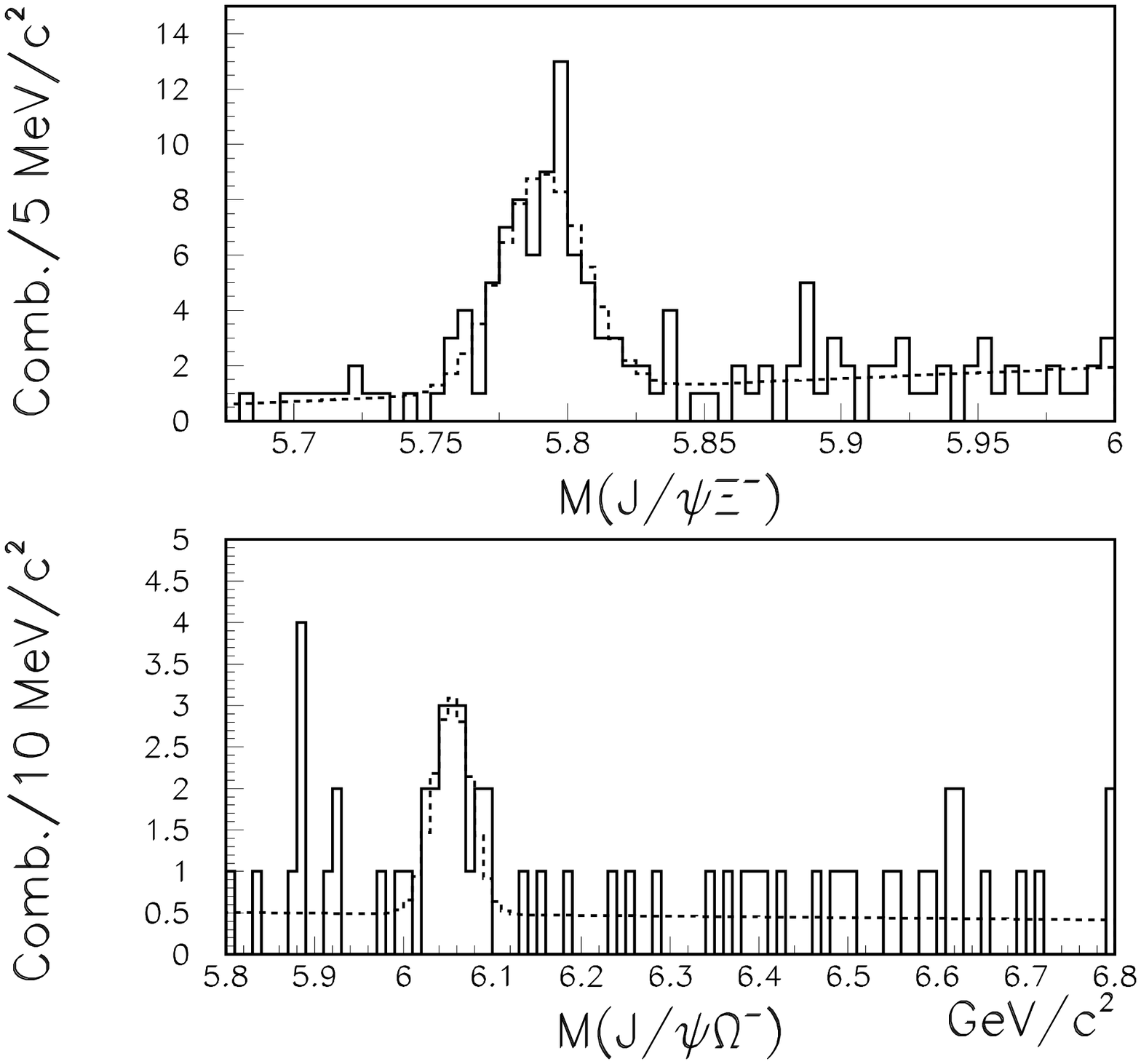}
\put(-410,150){\bf (a)}
\put(-285,120){\bf (b)}
\put(-35,150){\bf (c)}
\put(-35,65){\bf (d)}
\caption{Invariant mass distributions of 
(a) \Xib~and 
(b) \Omb~from D0, 
(c) \Xib~and 
(d) \Omb~from CDF.
}
\label{fig:baryons}
\end{figure}

Until 2006 only one bottom baryon, the \Lb, had been directly observed.
The $\Sb^{(*)}$~baryon states with quark content
$\Sb^{(*)+}=|buu\,\rangle$ and $\Sb^{(*)-}=|bdd\,\rangle$ have been
discovered by CDF~\cite{ref:CDF_sigmab} in 2007 through their strong
decay $\Sb^{(*)\pm} \ra \Lb\pi^{\pm}$ using fully reconstructed
$\Lb\ra\Lambda_c^+\pi^-$ candidates.  The \Xib~baryon with a quark
content of $\Xib=|bds\,\rangle$ was observed by CDF and D0 in the mode
$\Xib\ra J/\psi\Xi^-$ followed by $\Xi^-\ra\Lambda\pi^-$ with
$\Lambda\ra p\pi^-$ and $J/\psi\ra\mu^+\mu^-$.  The
D0~analysis~\cite{ref:D0_Xib} based on 1.3~fb$^{-1}$ of data finds
$(15.2\pm4.4^{+1.9}_{-0.4})$ \Xib~signal event with a Gaussian
significance of $5.2\,\sigma$ as shown in Figure~\ref{fig:baryons}(a).
D0 reports a mass of $m(\Xib)=(5774\pm11\pm15)~\mevcc$.  CDF observes
$(17.5\pm4.3)$ \Xib~signal events~\cite{ref:CDF_Xib} with a Gaussian
significance of $7.7\,\sigma$ and measures a \Xib~mass of
$m(\Xib)=(5792.9\pm2.5\pm1.7)~\mevcc$.

\subsection{Observation of  the \Omb~Baryon}

In August 2008, the D0 collaboration announced the observation of another
heavy bottom baryon~\cite{Abazov:2008qm}, the double strange $\Omb$
baryon with quark content $|bss\,\rangle$. Building on the previous
observation of the \Xib, D0 reconstructs $\Omb\ra J/\psi\Omega^-$
followed by $\Omega^-\ra\Lambda K^-$ in the same dataset using
1.3~fb$^{-1}$ of $p\bar p$~collisions.  A mass measurement of
$m(\Omb)=(6165\pm10\pm13)~\mevcc$ is reported based on an \Omb~signal of
$(17.8\pm4.9\pm0.8)$ events shown in Figure~\ref{fig:baryons}(b). The
significance of the observed signal is $5.4\,\sigma$ corresponding to a
probability of $6.7\times10^{-8}$ of it arising from background
fluctuation. D0 measures the \Omb~rate with respect to \Xib~production
to be $0.80\pm0.32^{+0.14}_{-0.22}$.

In May 2009, CDF released a comprehensive
reconstruction of bottom baryons with a $J/\psi$
in the final state~\cite{Aaltonen:2009ny}: 
$\Lb\ra J/\psi\Lambda$, $\Xib\ra J/\psi\Xi^-$ and
$\Omb\ra J/\psi\Omega^-$.  CDF reconstructs $(66^{+14}_{-9})$ \Xib\ and
$(16^{+6}_{-4})$ \Omb~candidates shown in Fig.~\ref{fig:baryons}(c) and
(d), respectively.  A $5.5\,\sigma$~significance for an \Omb~observation
is reported together with mass measurements of
$m(\Xib)=(5790.9\pm2.6\pm0.9)~\mevcc$ and
$m(\Omb)=(6054.4\pm6.8\pm0.9)~\mevcc$ in good agreement with
theoretical predictions for the \Omb~mass.  While there is agreement between
the \Xib~mass measurements, a significant mass difference of
$(111\pm12\pm14)~\mevcc$ exists between the \Omb~masses reported by CDF
and D0. CDF normalizes the observed \Omb\ and \Xib~rates to its
\Lb~production where both ratios correspond to a \Omb~rate of
$0.27\pm0.12\pm0.01$ normalized to \Xib~production. 


\end{document}